\documentclass[twocolumn,prl,showpacs,preprintnumbers,amsmath,amssymb]{revtex4}
\usepackage{color}
\usepackage{bm, graphicx, amsmath}
\newcommand{\two}[2]{\begin{array}{c}\\[-1.5em]\scriptstyle #1\\[-.5em] \scriptstyle #2\end{array}}

\begin{document}

\title{Relations between coherence and path information}
\author{Emilio Bagan$^{1,3}$, J\'anos A. Bergou$^{1,2}$, Seth S. Cottrell$^{4}$, and Mark Hillery$^{1,2}$}
\affiliation{$^{1}$Department of Physics and Astronomy, Hunter College of the City University of New York, 695 Park Avenue, New York, NY 10065 USA \\ 
$^{2}$Graduate Center of the City University of New York, 365 Fifth Avenue, New York, NY 10016 \\ 
$^{3}$F\'{i}sica Te\`{o}rica: Informaci\'{o} i Fen\`{o}mens Qu\`antics, Universitat Aut\`{o}noma de Barcelona, 08193 Bellaterra (Barcelona), Spain \\ 
$^{4}$Department of Mathematics, New York City College of Technology, 300 Jay Street, Brooklyn, NY 11201}

\begin{abstract}
We find two relations between coherence and path-information in a multi-path interferometer.  The first builds on earlier results for the two-path interferometer, which used minimum-error state discrimination between detector states to provide the path information.  For visibility, which was used in the two-path case, we substitute a recently defined $l_{1}$ measure of quantum coherence.  The second is an entropic relation in which the path information is characterized by the mutual information between the detector states and the outcome of the measurement performed on them, and the coherence measure is one based on relative entropy.
\end{abstract}


\maketitle

Recently a theory of quantum coherence as a resource for quantum information processing was proposed along with two possible coherence measures, an entropic measure and an $l_{1}$ measure \cite{baumgratz}.  This work has led to a renewed interest in the properties of quantum coherence~\cite{winter,hall}.  The $l_{1}$ measure is of interest, because it is, in principle, observable.  It depends on the magnitudes of off-diagonal density matrix elements, whose real and imaginary parts can be estimated.  The entropic measure is the difference between the von Neumann entropies of a density matrix and a diagonal density matrix formed from its diagonal elements.  In this paper we focus on the duality relations between these coherence measures and the which-path information for a particle going through an interferometer with two or more internal paths. 

It is well known that a particle going through an interferometer can exhibit wave or particle properties.  The particle properties are characterized by how much information one has about which path the particle took through the device.  The wave properties determine the visibility of the interference pattern.  There is an inverse relation between the particle and wave properties, the stronger one is the weaker is the other.  This was studied for interferometers with two internal paths in a quantitative way by Wootters and Zurek \cite{wootters}. The relation was put into an elegant form by Greenberger and YaSin  \cite{greenberger}
\begin{equation}
\label{dual}
D^{2}+V^{2}\leq 1,
\end{equation}
where $D$ is a measure of path information and $V$ is the visibility of the interference pattern. 
This work was carried further by Jaeger, \emph{et al}. \cite{jaeger}, who proposed a possible definition for path information for interferometers with more than two paths.  Wootters and Zurek employed a path detector in their analysis but did not derive a relation of the form given in Eq.\ (\ref{dual}).  In the subsequent work \cite{greenberger,jaeger}, which did derive a path-visibility relation, the path information is related not to information from a detector but to the preparation of the particle state, i.e.\ whether it is more likely to be in one path rather than in the other.  In a seminal study, Englert \cite{englert} combined these approaches.  He introduced detectors into the problem in order to define the path information and derived a relation between this type of path information and the visibility that took the form of Eq.\ (\ref{dual}).  In his model, a system of detectors, one in each path, is coupled to the paths, so that when the particle passes through the interferometer correlations are produced between the path states and the detector states.  Path information is then related to the distinguishability of the detector states.  If the detector states are orthogonal, one has perfect path information, but if they are not, then the information one can obtain about the path is smaller.

The first derivation of a path-visibility relation for more than two paths is due to D\"{u}rr \cite{durr}.  Expressing the density matrix of the particle inside the interferometer in a path basis, in which each path corresponds to one of a set of orthonormal states, his measure of path information depended on the diagonal elements of the density matrix, and his measure of visibility depended on the off-diagonal elements.  Some difficulties with the definition introduced by D\"{u}rr were pointed out in \cite{bimonte} and this, in turn, following the earlier discussion in \cite{englert2}, led to proper definitions of the quantities that are free of the difficulties~\cite{jakob} and also to some alternative definitions~\cite{englert3}.

We want to emphasize that in \cite{durr,bimonte,englert2,jakob,englert3} an $l_{2}$ measure of coherence was employed and, hence, the results presented there are relations between second moments that, in turn, are closely connected to uncertainty relations.  As argued in \cite{baumgratz}{\color{red},} a proper operational definition of coherence must be related to first moments, i.e., an $l_{1}$ measure. There they defined the $l_{1}$ coherence of a density matrix $\rho$ to be
\begin{equation}
C_{l_{1}}(\rho) =\sum_{\two{i,j=1}{i\neq j}}^{N} |\rho_{ij}|  .
\label{C}
\end{equation}
Here we shall consider a normalized version of this quantity given by $X=(1/N)C_{l_{1}}(\rho)$, which has the property that $0 \leq X \leq (N-1)/N$.  There is also an entropic measure of coherence that satisfies the criteria in \cite{baumgratz}, which, as we shall see, also leads to a duality relation.  Both definitions of coherence are basis dependent. We will treat the case of the $l_{1}$ measure first.

It is natural to search for a duality relation between path coherence and path information. The first approach to this problem using the~$l_{1}$ measure and a detector in each path was taken by Bera, \emph{et al.} \cite{pati}.  The discrimination of the detector states can be done in a number of ways, and the fact that the states are not orthogonal means it cannot be done perfectly.  Bera, \emph{et al}.\ used unambiguous discrimination, in which one never obtains a wrong answer, but the procedure can sometimes fail providing no information about the detector state.  They found that the sum of the path coherence and an upper bound to the probability of successfully discriminating the detector states is less than or equal to one.  This does not produce a relation of the form given in Eq.\ (\ref{dual}).  Furthermore, unambiguous discrimination is not possible when the detector states are linearly dependent. When Englert derived his relation, which did take the form of Eq.\ (\ref{dual}), he  used minimum-error state discrimination \cite{englert}, which is always possible, even if the detector states are linearly dependent.  In this procedure, one always obtains a result, but it can be wrong, though the probability of making an error is minimized. The  probability of successfully identifying the detector states, $P_{\rm s}$, quantifies the available path information via optimized measurements.

In this letter we study the duality in the $N$-path interferometer  between coherence and path information.  Our first result is a relation between these two quantities that has a form similar to Eq.~(\ref{dual}),
\begin{equation}
\label{main}
\left( P_{\rm s}-\frac{1}{N}\right)^{2} + X^{2} \leq \left(1-\frac{1}{N}\right)^{2}  .
\end{equation}
The reason $P_{s}-(1/N)$, rather than just $P_{s}$, appears is that it is the measure of how much better we can do by using prior information and detectors than by just guessing. With no prior information about the path and no detectors, we must assume that each path is equally likely. Then, if we just guess the path our probability of being right is $1/N$, which is the worst case scenario. If we read out the detectors and use prior information, our probability of being right is $P_{s}$. Note that if the detector states are orthogonal to each other, the two sides of the inequality are equal, so the inequality is tight.

In order to derive \eqref{main} we start with a particle entering an $N$-port interferometer via a generalized beamsplitter that puts it in the superposition state
\begin{equation}
|\psi\rangle =\sum_{i=1}^{N} \sqrt{p_i}|i\rangle .
\label{instate}
\end{equation}
The orthonormal basis states, $|i\rangle$, $i=1,2,\ldots, N$ correspond to the $N$ possible paths and span the $N$ dimensional Hilbert space, $\mathcal{H}_{p}$. Equation~\eqref{instate} represents the most general state of the particle inside the interferometer.  

While in the interferometer, the particle interacts with another system, called the detector.  The detector starts in a global state $|\eta_{0}\rangle$. The interaction of the particle with the detector is described by the controlled unitary $U(|i\rangle |\eta_{0}\rangle ) =  |i\rangle|\eta_{i}\rangle$, which entangles the path degree of freedom $|i\rangle$ of the particle with the detector state $|\eta_{i}\rangle$. After the particle has interacted with the detector, the state of the entire system is
\begin{equation}
|\Psi\rangle = \sum_{i=1}^{N}\sqrt{p_i} |i\rangle |\eta_{i}\rangle .
\end{equation}
Tracing out the detector, we find that the particle density matrix is given by
\begin{equation}
\rho={\rm Tr}_{\rm det} \left(|\Psi\rangle\langle\Psi|\right)=\sum_{i,j=1}^N\sqrt{p_{i}p_{j}} \langle\eta_{j}|\eta_{i}\rangle\, |i\rangle\langle j|, 
\label{rho}
 \end{equation}
 which, in turn, yields for our coherence measure $X$
\begin{equation}
X=\frac{1}{N}C_{l_{1}}(\rho) = \frac{1}{N} \sum_{\two{i,j=1}{i \neq j}}^{N} \sqrt{p_{i}p_{j}} |\langle \eta_{j}| \eta_{i}\rangle| .
\label{X}
\end{equation}

Since path information is encoded in the detector states we also need to introduce the detector density matrix, $\rho_{\rm det}$. Tracing out the particle states, we find
\begin{equation}
\rho_{\rm det} = {\rm Tr}_{\rm particle} \left(|\Psi\rangle\langle\Psi|\right) = \sum_{i=1}^N p_{i} \rho_{i}\, , 
\label{rhodet}
 \end{equation}
where $\rho_{i} =  |\eta_{i}\rangle \langle \eta_{i}|$. In order to obtain which-path information, we need to discriminate among the states $\{|\eta_{i}\rangle\}$.  To this end, we will employ the minimum-error strategy.  For~$N$ states, we have an $N$-element POVM with elements $\Pi_{i} \geq 0$,  which satisfy $\sum_{i=1}^{N}\Pi_{i}={\mathbb I}$.  The probability that if we are given the state $|\eta_{j}\rangle$ detector $i$ clicks is $\langle \eta_{j}|\Pi_{i}|\eta_{j}\rangle$.  We identify a click in detector $i$ with the detection of the state $|\eta_{i}\rangle$, so the average probability of successfully identifying the state is
\begin{equation}
P_{s}=\sum_{i=1}^{N}p_i  \langle \eta_{i}|\Pi_{i}|\eta_{i}\rangle=\sum_{i=1}^Np_i{\rm Tr}(\Pi_i\rho_i) .
\label{success}
\end{equation}
In minimum-error state discrimination, we seek to find a POVM that maximizes $P_{s}$.  The solution to the problem is known in complete generality for two states \cite{helstrom}, but only in special cases for more than two states.  Here we shall employ an upper bound on the success probability to obtain our main result.

There are several upper bounds on the success probability for minimum-error state discrimination \cite{qiu1,montanaro,tyson,qiu2,bae}. However, for our goals we find that another one, which we first state and later prove, is more useful.  If we have $N$ density matrices, $\{ \rho_{i}|j=1,2,\ldots N\}$, where $\rho_{i}$ appears with probability $p_{i}$, then the success probability for minimum-error state discrimination obeys
\begin{equation}
P_{\rm s} \le \frac{1}{N} + \frac{1}{2N}\sum_{i,j=1}^{N} \| \Lambda_{ij}\|_{1} ,
\label{upperbound}
\end{equation}
where $\Lambda_{ij}=p_i \rho_{i}-p_j \rho_{j}$ is the Helstrom matrix of the pair of states $\rho_i$, $\rho_j$, and the norm in this inequality is the trace norm.  In the case of pure states, $\rho_{i}=|\eta_{i}\rangle\langle\eta_{i}|$, we find, by diagonalizing the operator $\Lambda_{ij}=p_i |\eta_{i}\rangle\langle\eta_{i}| -p_j |\eta_{j}\rangle\langle\eta_{j}|$, that 
\begin{equation}
\|\Lambda_{ij}\|_{1} = 2\sqrt{ \left(\frac{p_i+p_j}{2}\right)^2- p_{i}p_{j}|\langle \eta_{i}| \eta_{j}\rangle |^{2}} .
\label{HelstromPure}
\end{equation}

This implies that the average probability of successfully identifying the detector state, entangled with a given path, is bounded above by Eq.~(\ref{upperbound}), with $ \| \Lambda_{ij}\|_{1} $ given by (\ref{HelstromPure}).
The quantity~$X$, which describes the coherence, is given in Eq.\ (\ref{X}).  This gives us the upper bound for the  expression on the left-hand side of Eq.\ (\ref{main}),
\begin{eqnarray}
\hspace{-3em}\left(P_{\rm s}\!-\!\frac{1}{N}\right)^2\!\!\!+\!X^2\! &\leq& \! \frac{1}{N^2}\!\!\!\! \sum_{\two{i,j=1}{i\neq j}}^{N} \!\!\!\sum_{\two{k,l=1}{k\neq l}}^{N}\! \!\!\!\Big(\frac{1}{4} \| \Lambda_{ij}\|_{1}  \| \Lambda_{kl}\|_{1} \! \nonumber\\
\hspace{-1em}
&+&\sqrt{p_{i}p_{j}}|\langle \eta_{i}| \eta_{j}\rangle | \sqrt{p_{k}p_{l}} |\langle \eta_{k}| \eta_{l}\rangle |\Big).
\label{bound1}
\end{eqnarray}
%

For fixed $i$ and $j$, the pair $( \frac{1}{2}\| \Lambda_{ij}\|_{1},\sqrt{p_{i}p_{j}}|\langle \eta_{i}| \eta_{j}\rangle |)$ can be viewed as a bra vector $\langle v_{ij}|$ of length $(p_i+p_j)/2$ (and, similarly, for fixed $k$ and $l$). The term in parentheses in the r.h.s. of \eqref{bound1} is the scalar product, $\langle v_{ij}|v_{kl} \rangle$, of two such vectors. Using the Schwarz inequality, the r.h.s. can be bounded above by
\begin{equation}
 \frac{1}{N^2}\Bigg(\!\! \sum_{\two{i,j=1}{j\neq k}}^{N} \!\!\!\frac{p_i+p_j}{2}\Bigg)^2=\left(1-\frac{1}{N}\right)^2 ,
\end{equation}
 and we recover Eq.~\eqref{main}.

We now give the proof of Eq. \eqref{upperbound}. The success probability of the $N$-element POVM was introduced in \eqref{success}.  An upper bound for the individual terms in the success probability can be found as  
\begin{eqnarray}
p_i{\rm Tr}(\Pi_{i}\rho_{i}) & = & p_j {\rm Tr}(\Pi_{i}\rho_{j})+{\rm Tr}\left(\Pi_i\Lambda_{ij}\right)\nonumber \\
&\leq &p_j{\rm Tr}(\Pi_{i}\rho_{j})+\max_{0 \le \Pi\le{\mathbb I}}{\rm Tr}\left(\Pi \Lambda_{ij}\right)\nonumber\\
&= &p_j{\rm Tr}(\Pi_{i}\rho_{j})+{\rm Tr}\left(\Lambda_{ij,+}\right)\nonumber\\
&= &p_j {\rm Tr}(\Pi_{i}\rho_{j}) + \frac{p_{i}\!-\!p_{j} \!+\!\| \Lambda_{ij}\|_{1}}{2}, 
\label{ineq}
\end{eqnarray}
where the subscript ``$+$'' stands for {\em positive part} of the operator, i.e.\ if $P_{+}$ is the projection onto the space of eigenvectors of the hermitian operator $\Lambda_{ij}$ with positive eigenvalues, then $\Lambda_{ij,+}= P_{+}\Lambda_{ij} P_{+}$.   Similarly, if $P_{-}$ is the projection onto the space of eigenvectors with non-positive eigenvalues, then we define $\Lambda_{ij,-}=P_{-}\Lambda_{ij} P_{-}$.  For the inequality in the second line, we have used the fact that for $\Lambda_{ij}$, one has ${\rm Tr}(\Pi_i \Lambda_{ij}) \le \max_{0\le\Pi\le{\mathbb I}}{\rm Tr}(\Pi\Lambda_{ij})$, since the maximization is over the set of  positive operators less than the identity, which contains $\Pi_i$. The equality in the third line results from choosing $\Pi$ to be the projector onto the positive part of $\Lambda_{ij}$ and noting that ${\rm Tr}(\Pi\Lambda_{ij}) \leq {\rm Tr}(\Pi\Lambda_{ij,+}) \leq {\rm Tr}(\Lambda_{ij,+})$ for a positive operator $\Pi$ with operator norm less than or equal to $1$.  The equality in the last line uses the fact that ${\rm Tr}(\Lambda_{ij} )= p_{i}-p_{j} = {\rm Tr}(\Lambda_{ij,+}) + {\rm Tr}(\Lambda_{ij,-})$, and $\|\Lambda_{ij}\|_{1}={\rm Tr}(\Lambda_{ij,+})- {\rm Tr}(\Lambda_{ij,-})$ from where ${\rm Tr}(\Lambda_{ij,+}) = (p_{i}-p_{j} + \|\Lambda_{ij}\|_{1})/2$ follows.  Taking now the sum over $i,j;i \neq j$ of both sides in the inequality \eqref{ineq}, we find
\begin{eqnarray}
(N-1)P_{\rm s}  & \leq & 1 - P_{\rm s} +  \frac{1}{2} \sum_{\two{j,k=1}{j\neq k}}^{N} \| \Lambda_{ij}\|_{1}  .
\label{proof}
\end{eqnarray}
Noting that $\|\Lambda_{ij}\|_{1} = 0$ for $i=j$, the sum in the last term can be extended to include the $i=j$ terms, immediately yielding Eq.~\eqref{upperbound}. We want to point out that for $N=2$, Eq.~\eqref{upperbound} is actually an equality, it reproduces the Helstrom bound. Our bound generalizes this for arbitrary $N$. The first term on the r.h.s. of \eqref{upperbound} would be the result of pure guessing, so the second term can be regarded as the gain provided by the measurement that takes into account the available prior information (probability with which an individual detector states  occurs and the overlaps of the states). 

It is also possible to derive a duality relation using the entropic definition of coherence.  The relative entropy coherence measure for a density matrix $\rho$ is given by
\begin{equation}
C_{\rm rel\,ent}(\rho )=S(\rho_{\rm diag})-S(\rho ) ,
\end{equation}
where $\rho_{\rm diag}$ is a diagonal density matrix in the specified basis whose diagonal elements are the same as those of~$\rho$, and $S$ denotes the von Neumann entropy, with the logarithms taken base $2$.  In our case the relevant density matrix is given by Eq.~(\ref{rho}). This gives us
\begin{equation}
C_{\rm rel\,ent}(\rho ) = H(\{ p_{j}\}) - S(\rho) ,
\end{equation}
where $H(\{ p_{j}\}) = -\sum_{j=1}^{N} p_{j}\log p_{j}$ is the Shannon entropy.

For path information we can consider the mutual information between the detector states labeling the paths and the results of probing them.  The detector density matrix was introduced in \eqref{rhodet}, so $\rho_i$ appears with a probability of $p_{i}$.  Let $D$ be a random variable corresponding to the choice of detector state; it takes the value $i\in \{ 1,2,\ldots N\}$, corresponding to $\rho_{i}$, with probability $p_{i}$.  We probe the detector states with a POVM, ${\cal M}=\{\Pi_{i}|i=1,\ldots N\}$ in order to identify them, and thereby identify the path.  Let the random variable corresponding to the measurement result be $M$.  It takes values in the set $\{ 1,2,\ldots N\}$, with $i$ corresponding to the detection of the state $\rho_{i}$. The joint distribution for the two variables is given by $p(M=i,D=j)= {\rm Tr}(\Pi_{i}\rho_{j})p_{j}$. Note that this situation is analogous to one in which Alice sends the state $\rho_i$ with a probability of $p_{i}$ to Bob, and Bob performs a state discrimination measurement in order to determine what state he received.  We will quantify the path information by the mutual information, $H(M\!:\!D)$.  If the random variables are perfectly correlated this is $H(D)=H(\{p_i\})$, while if they are uncorrelated this is equal to zero.

In this situation we can make use of the Holevo bound~\cite{nielsen}, which states that if Alice sends $\rho_{i}$ with probability $p_{i}$, and Bob measures the state he receives, then
\begin{equation}
H(M\!:\!D) \leq S(\rho_{\rm det}) -\sum_{i} p_{i} S(\rho_{i}) .
\end{equation}
We recall that in our case $\rho_{i}=|\eta_{i}\rangle\langle\eta_{i}|$, which means that the second term above is zero so that $H(M\!:\!D)\leq S(\rho_{\rm det})$.  Now $\rho$ and $\rho_{\rm det}$ are reduced density matrices of the same pure state [recall \eqref{rho} and \eqref{rhodet}], and, therefore, $S(\rho)=S(\rho_{\rm det})$.  Consequently, we have that
\begin{equation}
C_{\rm rel\,ent}(\rho) + H(M\!:\!D) \leq H(\{p_{i}\}) ,
\label{entropic bound}
\end{equation}
which is an entropic version of the coherence-path-information duality relation.  The relation is tight, because the bound is attained when the detector states are orthogonal.

Since the bound (\ref{entropic bound}) holds for any measurement, it also holds for the accessible information, defined as
${\rm Acc}(D)=\max_{\cal M} H(M\!:\!D)$, where the maximization is over all POVMs. Thus, we can also write
$$
C_{\rm rel\,ent}(\rho) + {\rm Acc}(D) \leq H(\{p_{i}\}).
$$
We also note that our bound based on the $l_1$ coherence measure holds for any deterministic discrimination protocol, for which all outcomes give a conclusive answer about the identity of the detector states, not just for the optimal measurement, with minimum discrimination error.

In summary, we have derived two relations relating the path information about a particle inside a multi-path interferometer to two recently defined measures of the coherence of a quantum system.  The first of these provides a generalization of the visibility-path-information relation derived by Englert for the two path case.  Previous studies used a number of different quantities as multi-path generalizations of the visibility, but our results here suggest that the recently defined $l_{1}$ and entropic coherence measures are strong candidates.

\emph{Acknowledgment}. This publication was made possible through the support of a Grant from the John Templeton Foundation. The opinions expressed in this publication are those of the authors and do not necessarily reflect the views of the John Templeton Foundation. Partial financial support by a Grant from PSC-CUNY is also gratefully acknowledged. The research of EB was additionally supported by the Spanish MICINN, through contract FIS2013-40627-P, the Generalitat de
Catalunya CIRIT, contract  2014SGR-966, and ERDF: European Regional Development Fund. EB also thanks Hunter College for the hospitality extended to him during his research stay.



\begin{thebibliography}{99}
\bibitem{baumgratz} T.\ Baumgratz, M.\ Cramer, and M.\ B.\ Plenio, \prl {\bf 113}, 140401 (2014).
\bibitem{winter} A.\ Winter and D.\ Yang, arXiv1501:07975 [quant.ph].
\bibitem{hall} S.\ Cheng and M.\ J.\ Hall, arXiv:1508.03240 [quant-ph].
\bibitem{wootters} W.\ K.\ Wootters and W.\ H.\ Zurek, \prd {\bf 19}, 473 (1979).
\bibitem{greenberger} D.\ M.\ Greenberger and A.\ YaSin, Phys.\ Lett.\ A {\bf 128}, 391 (1988).
\bibitem{jaeger} G.\ Jaeger, A.\ Shimony, and L.\ Vaidman, \pra {\bf 51}, 54 (1995).
\bibitem{englert} B.-G.\ Englert, \prl {\bf 77}, 2154 (1996).
\bibitem{durr} S.\ D\"{u}rr, \pra {\bf 64}, 042113 (2001).
\bibitem{bimonte} G.\ Bimonte and R.\ Musto, J.\ Phys.\ A {\bf 36}, 11481 (2003).
\bibitem{englert2} B.-G.\ Englert and J. A. Bergou, Optics Communications {\bf 179}, 337 (2000).
\bibitem{jakob} M.\ Jakob and J.\ A.\ Bergou, \pra {\bf 76}, 052107 (2007).
\bibitem{englert3} B.-G.\ Englert, D.\ Kaszlikowski, L.\ C.\ Kwek, and W.\ H.\ Chee, International Journal of Quantum Information {\bf 6}, 129 (2008).
\bibitem{pati} M.\ N.\ Bera, T.\ Qureshi, M.\ A.\ Siddiqui, and A.\ K.\ Pati, \pra {\bf 92}, 012118 (2015).
\bibitem{helstrom} C.\ W.\ Helstrom, \emph{Quantum Detection and Estimation Theory} (Academic Press, New York, 1976).
\bibitem{ban} M.\ Ban, K.\ Kurokawa, R.\ Momose, and O.\ Hirota, Int.\ J.\ Th.\ Phys.\ {\bf 36}, 1268 (1997).
\bibitem{hayashi} M.\ Hayashi, A.\ Kawachi, and H.\ Kobayashi, Quant.\ Inf.\ Comput.\ {\bf 8}, 0345 (2008).
\bibitem{qiu1} D.\ W.\ Qiu, \pra {\bf 77}, 012328 (2008).
\bibitem{montanaro} A.\ Montanaro, \emph{IEEE Information Theory Workshop, ITW08} (IEEE, Piscataway, NJ, 2008), p. 378.
\bibitem{tyson} J.\ Tyson, J.\ Math.\ Phys.\ {\bf 50}, 032106 (2009).
\bibitem{qiu2} D.\ W.\ Qiu and L.\ Li, \pra {\bf 81}, 042329 (2010).
\bibitem{bae}W.\ Y.\ Hwang and J.\ Bae, J.\ Math.\ Phys.\ {\bf 51}, 022202 (2010).
\bibitem{nielsen} M.\ Nielsen and I.\ Chuang, \emph{Quantum Computation and Quantum Information} (Cambridge University Press, Cambridge, 2000).
\end{thebibliography}
\end{document}